\documentclass{ws-ijmpaar}
\usepackage[super,compress]{cite}
\usepackage{graphicx}

\begin{document}
\markboth{G. Feofilov, V. Kovalenko, A. Puchkov}{Correlation of strange particles production with multiplicity in a multi-pomeron exchange model}

%\catchline{}{}{}{}{}
%
%%%%%%%%%%%%%%%%%%%%%%%%%%%%%%%%%%%%%%%%%%%%%%%%%%%%%%%%%%%%%%%%%%%%

\title{Correlations with strange particles in a multi-pomeron exchange model}

\author{Grigory Feofilov, Vladimir Kovalenko$^\text{1}$, Andrei Puchkov}

\address{Saint Petersburg State University, 7/9 Universitetskaya nab., \\
St.Petersburg, 199034, Russia\\
$^\text{1}$v.kovalenko@spbu.ru}

\maketitle

\begin{history}
%\received{Day Month Year}
%\revised{Day Month Year}
\end{history}

\begin{abstract}
 We propose a further development of a multi-pomeron exchange model of pp and p$\bar p$ collisions at high energy. The model  describes consistently multiplicity, transverse momentum of charged particles and their correlation in wide energy range (from ISR to LHC) by introducing string collectivity effects in pp collisions. The particles species differentiation is implemented according to Schwinger mechanism. Due to the influence of higher resonances on the yields of observed particles, we implement the production of a large set of hadron resonances with cascade decays. We present the results on pion, kaon and proton yields as a function of the collision energy and the correlation of strange and multi-strange hadrons with multiplicity and compare the results with experimental data.

\end{abstract}

\ccode{PACS numbers: 13.85.-t, 14.40.Aq, 13.60.Rj, 13.30.Eg}

%\tableofcontents

\section{Introduction}	

The multi-pomeron exchange model [1-3] describes consistently multiplicitiy, transverse momentum and their correlation in wide energy range (from ISR to LHC) by introducing string
collectivity effects in pp collisions. The particles discrimination is implemented according to
Schwinger mechanism [4].
In this model it is assumed for simplicity that all the produced particles have the same mass.
This approach can be justified at relatively low energies at which the spectra of the produced particles is dominated by $\pi$-mesons. However, at the LHC energies kaons and heavier
hadrons contribution is more than 15\%.
 At the same time, most of pions are producing
as a result of decay of the $\rho$-mesons and other resonances [5].
For the description of production of strange particles the  multi-pomeron exchange model was extended [1-3] by incorporation of four types of hadrons $\pi, K, \rho$-mesons and protons. 

In this paper we further develop the effective multipole exchange model, by taking into account the production of a wide range of hadrons, including resonances, in the mass range from 135 MeV to 2.5 GeV. The final hadron spectrum is formed taking into account  the cascade decays of resonances by strong interaction with the aid of an effective branching matrix.

Within the framework of the new model, the multiplicities of different types of particles were calculated and compared with the results of the experiment.

\section{Model description}

{{If the pp-collision spectrum contains several types of particles, then the multiparticle distribution of multiplicity and the transverse momentum of the prompt hadrons should be a linear combination of the corresponding one-particle functions: }}

\begin{eqnarray}
    \rho ( N_{ch}, p_{t} ) =
    C_{w}
    \sum
    \limits_{\nu}
    {\rho ( N_{ch}, p_{t}; \nu ) } = \ \ \ \ \ \ \  \ \ \ \ \ \ \ \\
    =C_{w}
    \sum
    \limits_{\nu}
    \sum
    \limits_{n=1}^{\infty}
    w_{n}(z) P(n, N_{ch}) g_{\nu}
    K (n)  \exp{
    \left(
    -
    \dfrac{\pi (p_{t}^2 + m_{\nu}^2) }
    {n^{\beta}t}
    \right)
    }\,,
   \label{maineq}
\end{eqnarray} 

{
\ \ \ \ \ \ \ \ \ \ \ \ \ \ \ \ \ \ \ \ \newline
Here}{ }
${{w}}_{{n}}\left({z}\right)${
is the probability of produciton of $n$ pomerons in an event,
${P}\text{ }\text{(}{n},{{N}}_{\text{ch}}\text{)
}\text{–}${ probability of the emission of }
${{N}}_{\text{ch}}${ charged particles from
} ${n}${ as a result of hadronization. }{$\delta
$}{
}{–}{ acceptance, the width of the (pseudo)-rapidity interval in which the particles are emitted.} $K(n)$ is a normalization factor [6].

\medskip

According to multi-pomeron model [2, 3], in the formula (\ref{maineq}):

\begin{equation}
  \label{a2}
w_{n}(z) =
    \dfrac{1}{n}
    \left(
    1 - \exp(-z)
    \sum
    \limits_{l=0}^{n-1}
    {\dfrac{z^l}{l!}}
    \right)\,,
    \;\;\;\;\;
   z =
   \dfrac{
   2 C \gamma s^{\Delta}
   }
   {
   R_{0}^{2} + \alpha^{'}\ln{(s)}
   }\,.
\end{equation}
\begin{equation}
   \label{a3}
P(n, N_{ch}) =
 \exp(-2nk \delta)
    \dfrac
    {(2nk\delta)^{N_{ch}}}
    {N_{ch}!}\,.
\end{equation}
The Regge parameters used in the calculations are the following [2,3]:
\begin{equation}
  \label{a4}
  \Delta =
  0.139\,,\;\alpha^{'}=0.21 \mathrm{ GeV}^{-2}\,,\;\gamma=1.77 \mathrm{ GeV}^{-2}\,,\;
  R_{0}^{2}=3.18 \mathrm{ GeV}^{-2}\,,\;C=1.5\,.
\end{equation}

{
{The probability of the production of a primary hadron of the type}
$\text{$\nu $}${ with transverse momentum }
${p}_{t}${, in accordance with the modified Schwinger mechanism [1,4], is proportional to the value}}

\begin{equation*}
{{g}}_{\text{$\nu $}}\text{exp}\left(-\frac{\text{$\pi
$}\left({{{p}}_{t}}^{2}+{{{m}}_{\nu
}}^{2}\right)}{{{n}}^{\text{$\beta $}}{t}}\right),
\end{equation*}
%\bigskip
{where parameter
${{m}}_{\text{$\nu $}}${ is a mass of } $\text{$\nu $}${ hadron and } $\beta
${ this parameter is responsible for the collectivity (effectively takes into account the fusion of strings). Index }
$\text{$\nu $}=\text{1},\text{2},\text{3}${
is from 1 to 380, with lowest value corresponding to light hardons ($\text{$\nu $}=\text{380}${ corresponds to } ${\pi
}^{0}${ mesons). The multiplier }
${{g}}_{\text{$\nu $}}${ takes into account the symmetry spin factor: }
${{g}}_{\text{$\nu $}}=2{S}_{\nu }+1${.}}

String collectivity parameter   $\beta $,
responsible for string collectivity, in the collision energy dependence is fixed proceeding from the description of ${\left\langle
{p}_{t}\right\rangle
}_{{N}_{\text{ch}}}-{N}_{\text{ch}}$ correlations in pp and $\text{p}\bar{{\text{p}}}$ collisions.
The procedure for fixing the parameters is in detail described in the works [2, 3].

\bigskip

{
It is important that the final hadron spectrum is significantly affected by the decay of resonances by strong interaction. In order to take it into account, we introduce the concept of an effective branching matrix ${M}_{\mu \nu
}${. The elements of this matrix are the average multiplicity of hadrons of the type } $\nu${, which are formed as a result of the decay of a hadron of the type} $\mu
${ by strong interaction (taking into account cascade decays).}}

{
{Elements of this matrix were calculated using the particle decay algorithm built into the THERMINATOR 2 Monte Carlo generator [7].}}

{
{Thus, the relative yields of particles corrected for the decays of hadrons have the form:}}
\begin{equation*}
{Y}_{\nu }{\sim}\sum _{\mu }{{M}_{\mu \nu }}{\cdot}\left(2{S}_{\mu
}+1\right){\cdot}\exp \left(-\frac{\pi \left({p}_{t}^{2}+{m}_{\mu
}^{2}\right)}{{n}^{\beta }t}\right),
\end{equation*}
{
where  ${S}_{\mu }$ is a spin of a particle of the type  $\mu $,  ${M}_{\mu \nu }$ is effective branching matrix.}
{
The yields of particles are normalized taking into account the conservation of total multiplicity ${N}_{\text{ch}}$.}

\section{Results}
\begin{figure}[b]
\centering
\includegraphics[width=8cm,clip]{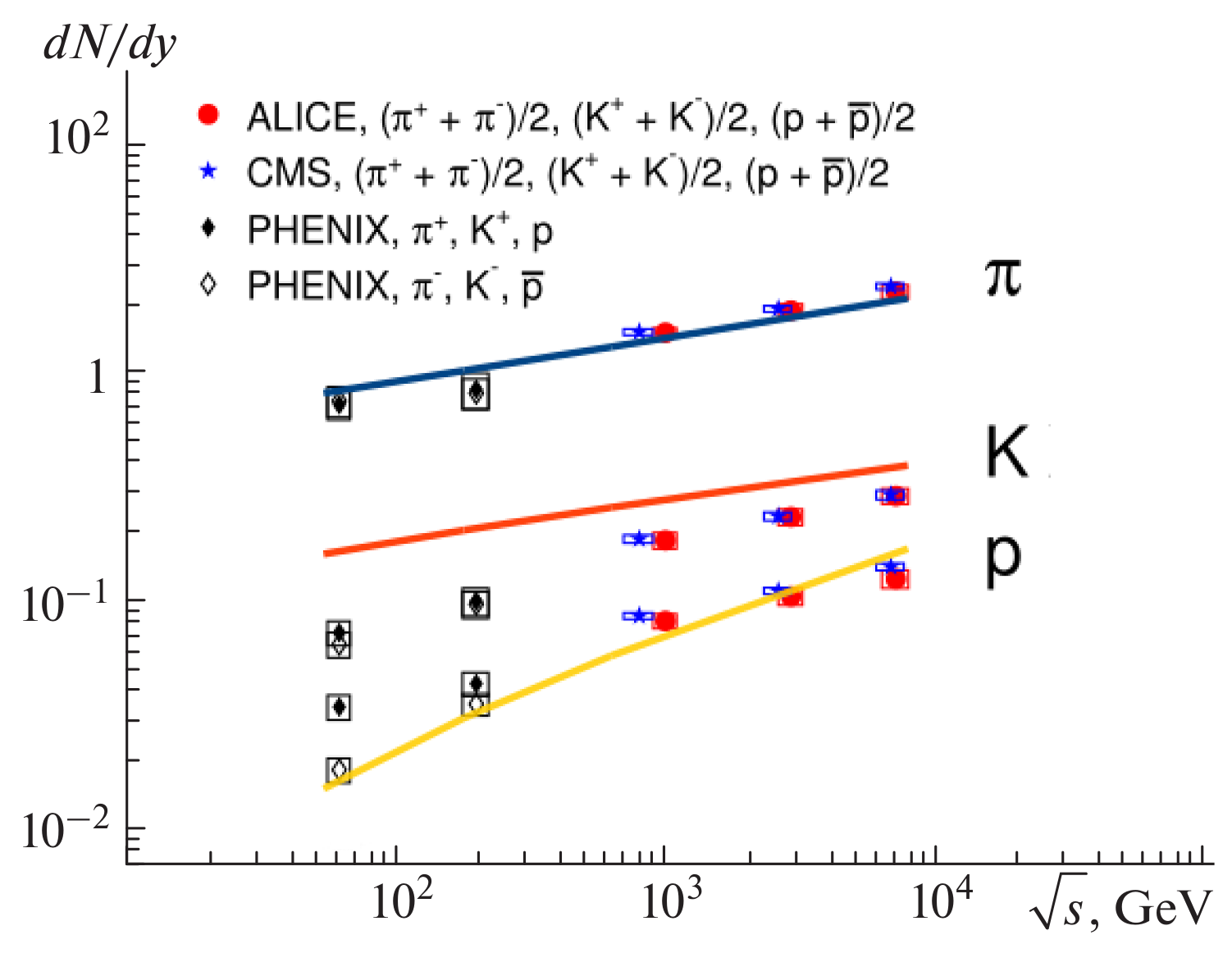}
\caption{Energy dependence of the mean multiplicity per rapidity for charged pions, kaons, and protons in pp collisions. The lines correspond to model calculations taking into account the resonances decays. The points denote experimental data (see references in [9]).}
\label{fig1}       % Give a unique label
\end{figure}

In Fig. 1 we present the results of calculations in the multipopomeron exchange model for the average multiplicity of charged pions, kaons and protons in pp collisions as a function of energy, and the corresponding experimental data (see references in [8]). It can be seen that a complete account of the decays of resonances significantly improves the agreement between the predictions of the model and experiment, compared with the more simplified model [6], especially at LHC energies. The remaining discrepancy in the SPS energy range can be explained by the influence of a nonzero baryochemical potential at such energies, which increases the relative fraction of protons and decreases the one of $K$ mesons.

Within the framework of this multi-pomeron model, the yields of strange and multi-strange hadrons in pp collisions at 7 TeV, as a function of the multiplicity of charged particles, were also calculated.
In a recent article of ALICE [9] it was shown that with the multiplicity of proton-proton collisions, the yield of strange and multi-strange particles, normalized to the multiplicity of pions, grows strongly. At the same time, the experimental data were poorly described by the available models. The best agreement was demonstrated by taking into account the increase in string tension with multiplicity.
It should be noted that the increased growth of strangeness was predicted for a long time ago in the string fusion model [10-12]. In the present model, the effects of string collectivity are effectively taken into account, so one can expect that similar strangeness enhancement can appear.

\begin{figure}[b]
\centering
\includegraphics[height=7cm,clip]{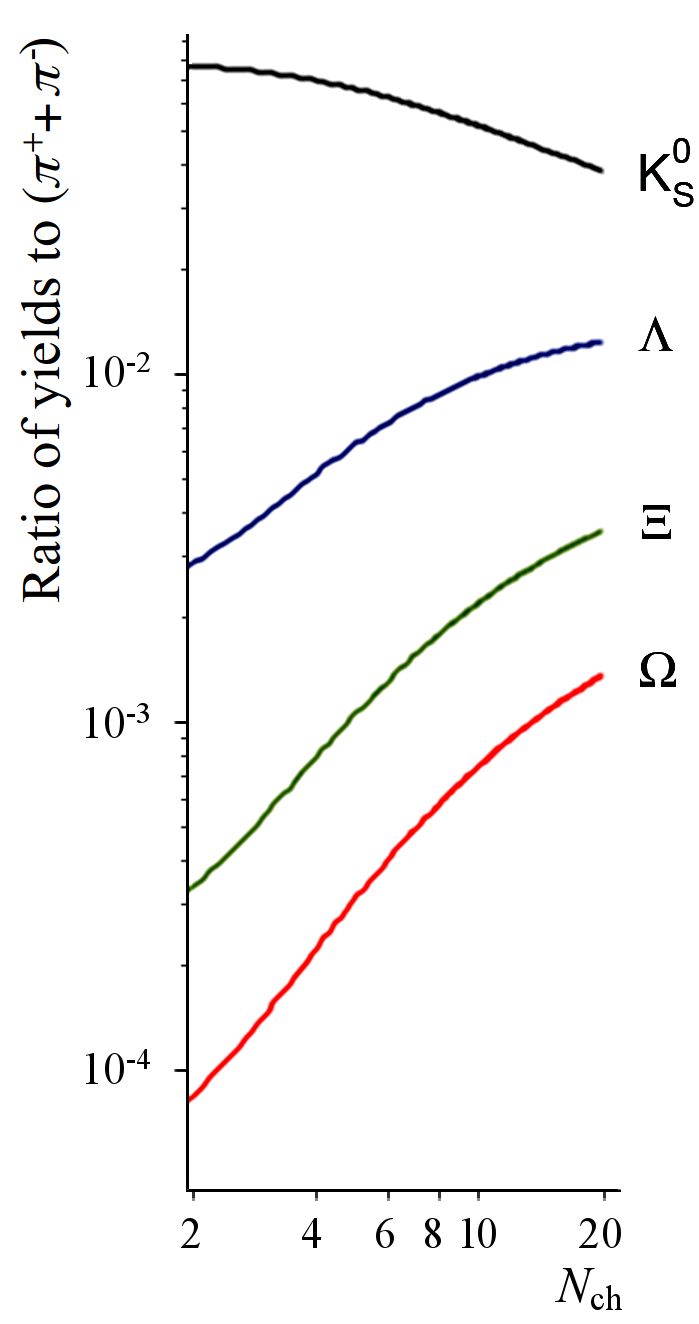} \hspace{0.8cm}
\includegraphics[height=7cm,clip]{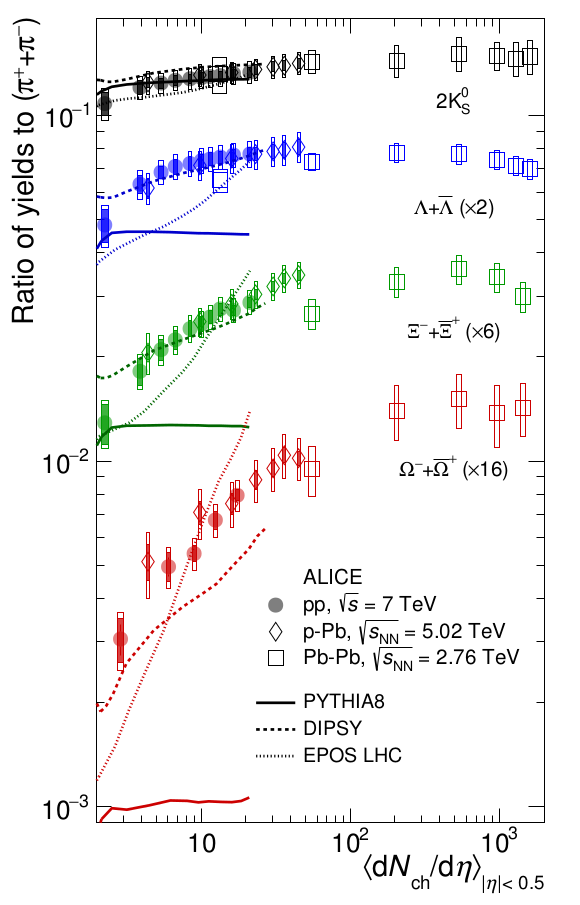}
\caption{Multiplicity dependence of the strange and multi-strange particle yield divided by charged pion multiplicity in pp-collisions at $\sqrt{s}=$7 TeV. Left plot -- model calculation, right -- experimental data [9]).}
\label{fig2}       % Give a unique label
\vspace{0.6cm}
\end{figure}

Fig. 2 shows the yields of baryons and mesons, normalized to the number of mesons, as a function of the charged multiplicity. It can be seen that the multi-pomeron model qualitatively correctly describes the rapid growth of the multiplicity of strange and multi-strange baryons. The differences between the model predictions for mesons and experiment appear to be related to a number of unaccounted factors, such as nontermal rescattering, the effects of preserving total strangeness, etc., this question requires further study.

\begin{figure}[h]
\centering
\includegraphics[width=5.4cm,clip]{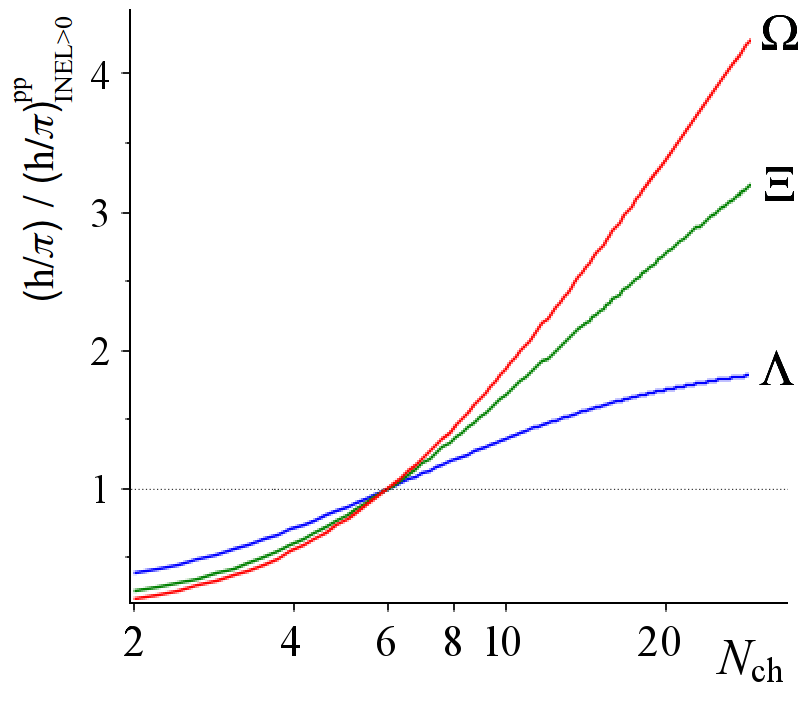}
\hspace{0.25cm}
\includegraphics[width=5.8cm,clip]{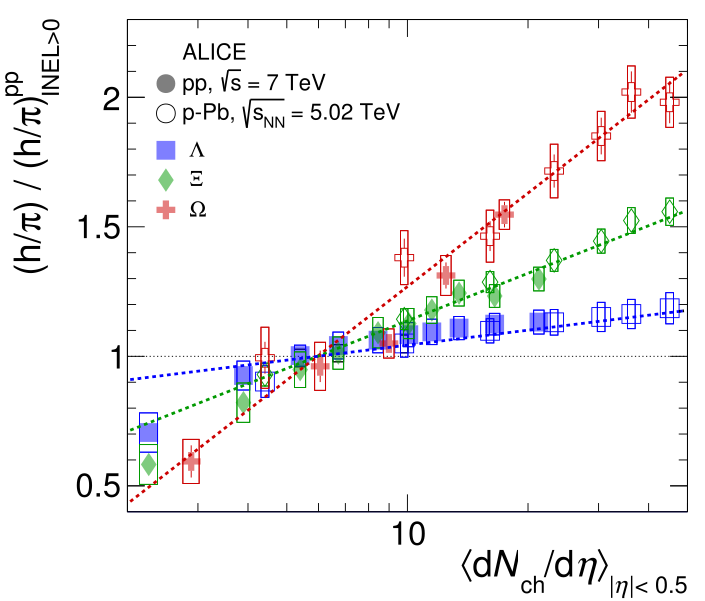}

\caption{Multiplicity dependence of the double ratio of multi-strange yield over charged pion yield, normalized by this value for minimum bias pp-collisions at $\sqrt{s}=$7 TeV.  Left plot shows the model calculation, right -- experimental data  [9]).}
\label{fig3}       % Give a unique label
\end{figure}

In Fig. 3 we show the relative yield of strange and multi-strange baryons, divided by the pion multiplicity, normalized to the value of this value in all pp collisions (without selection by multiplicity). It can be seen that the multi-pomeron model demonstrates an even stronger relative yield of strangeness, as in the experiment. This may be due to the fact that in this model the collective parameter effectively takes into account all the collective processes leading to a correlation between the multiplicity and the transverse momentum, although apparently there may be other contributions not related to the modification of the string tension [13].

It is worth noting that all the new strange results in this multi-pomeron model were obtained without any additional adjustment of the parameters. Additional adjustment of these parameters, taking into account the widened range of observed values, will improve the agreement of the results with experiment.
To summarize, comparison of the prediction data with the recent experimental data of ALICE can be considered as another confirmation of the effect of string tension growth with multiplicity (with string density). This phenomenon was investigated in the framework of the  PYTHIA Monte Carlo generator, and the dedicated tune for taking this effect into account was prepared [14].

%For tables use syntax in table~\ref{tab-1}.
%\begin{table}
%\centering
%\label{tab-1}       % Give a unique label
%% For LaTeX tables you can use
%\begin{tabular}{lll}
%\hline
%first & second & third  \\\hline
%number & number & number \\\hline
%\end{tabular}
%% Or use
%\vspace*{5cm}  % with the correct table height
%\end{table}

\section{Conclusions}

The further development of the effective model of multipole exchange intended for pp and p$\bar{\text{p}}$ collisions a wide energy range, which effectively takes into account the interaction between several strings. A correct generalization of this model to the case of the production of various types of particles is proposed. The yields of primary hadrons (including resonances) are calculated, and a correction is made for their decays with the aid of an effective branching matrix. It is shown that this approach significantly improves the agreement with the experiment of the $\pi$, $K$, $p$ yields in a wide energy range. It is demonstrated that the model qualitatively describes the strong growth of strange and multi-strange baryons with multiplicity in pp collisions at 7~TeV, recently discovered in the ALICE experiment.

\section{Acknowledgements}
The authors acknowledge Saint-Petersburg State University for a research grant 11.38.242.2015. 
%
%

%\begin{thebibliography}{000} %for 3 digits
%\begin{thebibliography}{00}  %for 2 digits

\end{document}